\documentclass[journal]{IEEEtran}
\usepackage{amsmath,amsfonts}
\usepackage{algorithmic}
\usepackage{algorithm}
\usepackage{array}
\usepackage[caption=false,font=normalsize,labelfont=sf,textfont=sf]{subfig}
\usepackage{textcomp}
\usepackage{stfloats}
\usepackage{url}
\usepackage{verbatim}
\usepackage{graphicx}
\usepackage{cite}
\usepackage{tikz}
\usepackage{textcomp}
\usepackage{changepage}
\usepackage{booktabs}
\usepackage{bm}
\usepackage{tabularx}

\begin{document}

\title{Preventing Adversarial AI Attacks Against Autonomous Situational Awareness: A Maritime Case Study}

\author{Mathew J. Walter,
        Aaron Barrett,
        and Kimberly Tam
\thanks{Mathew J. Walter is with The School of Engineering, Computing and Mathematics, University of Plymouth, Plymouth, UK, PL4 8AA (Email: mathew.walter@plymouth.ac.uk).}
\thanks{Aaron Barrett is with The School of Engineering, Computing and Mathematics, University of Plymouth, Plymouth, UK, PL4 8AA.}
\thanks{Kimberly Tam is with The School of Engineering, Computing and Mathematics, University of Plymouth, Plymouth, UK, PL4 8AA, and The Alan Turing Institute, British Library, London, UK, NW1 2DB.}
\thanks{Manuscript received November, 2024; revised August XX, 2025.}}




\maketitle

\begin{abstract}
Adversarial artificial intelligence (AI) attacks pose a significant threat to autonomous transportation, such as maritime vessels, that rely on AI components. Malicious actors can exploit these systems to deceive and manipulate AI-driven operations. This paper addresses three critical research challenges associated with adversarial AI: the limited scope of traditional defences, inadequate security metrics, and the need to build resilience beyond model-level defences. To address these challenges, we propose building defences utilising multiple inputs and data fusion to create defensive components and an AI security metric as a novel approach toward developing more secure AI systems. We name this approach the \texttt{Data Fusion Cyber Resilience} (DFCR) method, and we evaluate it through real-world demonstrations and comprehensive quantitative analyses, comparing a system built with the DFCR method against single-input models and models utilising existing state-of-the-art defences. The findings show that the DFCR approach significantly enhances resilience against adversarial machine learning attacks in maritime autonomous system operations, achieving up to a 35\% reduction in loss for successful multi-pronged perturbation attacks, up to a 100\% reduction in loss for successful adversarial patch attacks and up to 100\% reduction in loss for successful spoofing attacks when using these more resilient systems. We demonstrate how DFCR and DFCR confidence scores can reduce adversarial AI contact confidence and improve decision-making by the system, even when typical adversarial defences have been compromised. Ultimately, this work contributes to the development of more secure and resilient AI-driven systems against adversarial attacks.
\end{abstract}


\section{Introduction}

\IEEEPARstart{A}{rtificial} intelligence (AI) is rapidly permeating various aspects of our lives, offering significant benefits through task automation. This includes automating cyber-physical systems, such as transportation and industrial operations. The maritime sector is one of several domains embracing AI to capitalise on many benefits, ensuring organisations remain competitive. The International Maritime Organisation (IMO) categorises autonomy into four degrees, with the highest levels being degrees three and four. The proposed benefits of higher degrees of autonomy include significant operational benefits such as reduced crew and greater payload capacity, military utilisation in dangerous, contested, or Global Navigation Satellite System (GNSS) degraded/denied environments, greater automated decision making as well as increased safety and social benefits \cite{askari2022towards, porathe2014situation, morris2017worlds, ziajka2021costs, munim2019autonomous, kretschmann2017analyzing, felski2020ocean, tsvetkova2022creating}. 

Whilst AI can provide significant operational benefits, current research shows that AI models can harbour a significant number of vulnerabilities unique to AI systems and processes if they are not developed to be resilient. The terms adversarial AI (AAI) and adversarial machine learning (AML) were coined to describe these vulnerabilities \cite{goodfellow2014explaining, szegedy2013intriguing}. Organisations have acknowledged this threat by formulating measures such as OWASP's machine learning vulnerabilities top 10, NIST's AI Risk Management Framework (AI RMF), and MITRE's Adversarial Threat Landscape for Artificial-Intelligence Systems (ATLAS) threat modelling. Globally, government organisations, such as the United Kingdom's National Cyber Security Centre (NCSC), contributed to this cause with the ``Guidelines for Secure AI System Development'' in 2023 \cite{NCSC2023}. A significant emphasis was placed on adopting a secure-by-design development approach \cite{NCSC2022, NCSC2023}.

Many of the AAI concerns have already materialised within the domains of AAI and explainable AI (XAI) \cite{wolf2017we,grosse2023machine}, including adversarial attacks against autonomous vehicles \cite{qayyum2020securing, girdhar2023cybersecurity, joshi2024artificial}. Furthermore, attacks against AI within critical national infrastructure (CNI) and transportation have the potential for devastating consequences, resulting in a significant loss of money, reputation and life \cite{tam2023quantifying}. Such threats become increasingly likely with the exponential-like uptake in AI, greater reliance on AI for critical decision-making, and more effective AAI methods expanding the threat landscape.

The defence and resilience of AI systems remains an underdeveloped field with numerous key challenges. The limitations of AAI defences can be grouped into three main categories, which we aim to address in this work:
\begin{enumerate}
\item \textbf{Limited Scope of Traditional Defences}: Traditional AAI defences are often restricted to countering a single type of attack. They can lack consistent accuracy, and many operate effectively within limited or restricted conditions. Therefore, we explore the development of defences that are effective across multiple attack types. 
\item\textbf{Inadequate Security Metrics}: Existing metrics, such as model confidence, offer limited insight into attacks and are insufficient for integrating security into the system's decision-making process. Existing metrics to measure and understand risks from AAI are very limited. We emphasise the need for security and robustness metrics, 
such as the security-inclusive confidence score proposed in this paper. 
\item\textbf{Resilience Beyond Model Defences}: Existing traditional defences do not consider resilience, the ability to continue functioning during an attack, which is critical for autonomy. By adopting a defence-in-depth approach, we investigate whether it is possible to create a more robust system to mitigate the effects of attacks — even if the model's defences are bypassed. 
\end{enumerate}

This work proposes and evaluates a novel approach, the Data Fusion Cyber Resilience (DFCR) method, to build more secure AI systems by using multiple input sources, data fusion methods, and defence-oriented components tailored to the specific application and environment to address these challenges. This approach enhances system security and resilience, effectively overcoming the aforementioned limitations compared to single AAI defence methods such as input image compression or adversarial training. Moreover, we demonstrate how the proposed method can provide defence over a range of attacks, rather than being limited to mitigating a single type of attack, unlike most AAI defence methods. It can also be utilised to generate metrics which incorporate system security and develop more resilient AI systems. In this paper, we emphasise an important terminology distinction between AI models and AI systems. AI models refer specifically to standalone models, while AI systems incorporate the model as part of a broader framework, including processes such as data preprocessing, feature extraction, model defences and post-processing.

We measured the impact of the DFCR method in two ways. First, we conducted sea trials for both AAI and AAI defences to evaluate their real-world practicality. This was a critical aspect of the study, as previous research \cite{walter2023adversarial, walter2024red} highlighted that evaluations conducted in low-entropy laboratory settings often exhibit different behaviours when applied in the complex and dynamic conditions of real-world environments. To enhance the realism of the evaluations, we employed maritime autonomous systems (MAS) during these trials. Since real-world environments are the ultimate intended operational domain for AI tools, to better understand the actual effects of attacks and defences, this study emphasised evaluating defences in situ. Sea trials enabled one to consider and compare attacks on both operational and theoretical levels, providing insights into the limitations and practicality of the methods under real-world conditions and revealing notable disparities between laboratory-based and in situ AAI research. The second method of impact measurement quantitatively evaluated the DFCR approach by comparing it against existing state-of-the-art defences and single-input models to assess the attack success rate.

Through these methods of assessing impact, we are able to evaluate and demonstrate the paper's novel contributions:
\begin{enumerate}
    \item Building defences utilising multiple inputs and data fusion to create defensive components (DFCR), and a novel AI security metric.
    \item Using real-world data collected by MAS at sea, show how effective the system and metric are against AML attacks for MAS operations (e.g., object detection).
    \item Comprehensive quantitative evaluation of the security-accuracy trade-off of the DFCR approach against non-secure and single-input models and existing state-of-the-art defences.
\end{enumerate}

The remainder of this paper is structured as follows. In Section~\ref{sec:literature}, we review the relevant literature regarding maritime data fusion, AI security, and maritime AI security. Section~\ref{sec:modelArch} contains the methodology implemented for creating the DFRC system and the DFCR security metric. The experimental setup and equipment details are contained in Section~\ref{sec:method}. The results and analysis are highlighted in Section~\ref{sec:results}. Finally, we discuss future work in Section~\ref{sec:discuss} and provide a conclusion in Section~\ref{sec:Conclusion}.

\section{Existing Background }
\label{sec:literature}

\subsubsection{AI Security Overview}

In the early works of \cite{kearns1988learning,dalvi2004adversarial,lowd2005adversarial}, adversarial attacks were first introduced against spam filters. Significant attention was raised when \cite{szegedy2013intriguing} showed how computer vision neural networks (convolutional neural networks) were vulnerable to adversarial examples and introduced the Large-BFGS method to create adversarial perturbations. Biggio et al. \cite{biggio2013evasion} was also a key author in the initial exploration of neural network vulnerabilities. The work of \cite{goodfellow2014explaining} formulated the fast gradient sign method (FGSM) to attack computer vision models with open-box (white-box) gradient-based attacks. In \cite{kurakin2016adversarial}, FGSM was adapted to create three new variants; these included the One-step Target Class method to optimise the adversarial example toward a particular class, the Basic Iterative Method (BIM), which could generate multiple examples via an iterative method, and the Iterative Least-likely Class Method which iteratively perturbed the adversarial example toward the weakest recognised class. 

Papernot et al. \cite{papernot2016limitations} proposed the Jacobian saliency maps attack (JSMA), which utilised the Jacobian of a model to perturb the solution toward a desired output (i.e., how a pixel change affects the predicted output). Papernot et al. also proposed a method to find a sensitivity direction by using the Jacobian matrix of the model. Similarly, \cite{su2019one} showed an attack method which only required the change of one pixel in the image. The work of \cite{carlini2017towards} proposed a method to minimise the loss between the target function and three norms $L_{0}, L_{2}, L_{\inf}$ between the adversarial example and the original image. The work of \cite{madry2017towards} utilised projected gradient descent (PGD) to minimise a loss function and project the adversarial example into the space of legal solutions. Deepfool was proposed in \cite{moosavi2016deepfool}, which created untargeted adversarial examples within a $L_{2}$ norm.

Non-evasion attacks include poisoning-based attacks \cite{barreno2006can, biggio2011support, gu2017badnets} and privacy-based attacks, e.g., model inversion attacks against APIs \cite{frederickson2018attack}, property inference \cite{ateniese2015hacking}, membership inference \cite{shokri2017membership, homer2008resolving} and model extraction attacks \cite{tramer2016stealing}.

Transformer security has gained significant attention, especially adversarial attacks on Large Language Models (LLMs), including poisoning, prompt injections, Denial-of-service (DoS), jailbreaking, data extraction, and membership inference \cite{shayegani2023survey, yao2024survey, chowdhury2024breaking}. Studies  \cite{bhojanapalli2021understanding, aldahdooh2021reveal} also suggest that Vision Transformers (ViTs) may be more robust than convolutional neural networks (CNNs) in tasks like object detection and classification, as their self-attention mechanism captures global features, enhancing resistance to noise and adversarial attacks. However, \cite{fu2022patch} finds ViTs can still be vulnerable under certain conditions (global feature perturbation) using specific transformer-based attacks, though generally more robust to existing attacks. Recent research highlights energy-focused attacks on ViTs. For example, \cite{wei2022towards, wei2023towards} introduce ``Pay No Attention" (PNA) and ``PatchOut" attacks, which enhance transferability and diversity in adversarial approaches for ViTs. Additionally, \cite{navaneet2023slowformer} presents ``SlowFormer", a universal patch that increases computational load and energy consumption. Similarly, \cite{yehezkel2024desparsify} describes the ``DeSparsify Attack" targeting ViTs with token sparsification methods (e.g., ATS, AdaViT, A-ViT) to raise computational demands without disrupting classification.

\subsubsection{Maritime AI Security}
There have been few academic papers regarding maritime AI security compared to more established AI topics, with most released in recent years. This indicates that this is a novel area of research, and also a quickly developing one within maritime cyber security research \cite{vineetha2024literature}. AI cyber security or resilience for MAS is becoming increasingly important with the increasing use of AI in MAS and into the future. 

The work of \cite{yoo2022artificial} considers potential attacks on future AI maritime autonomous vessels, whereas \cite{walter2023adversarial} showcased some of the first preliminary adversarial AI test cases/attacks against MAS. Other works, including \cite{lee2023vulnerability}, propose poisoning-based adversarial AI attacks against MAS. Adversarial waypoint injection attacks against MAS were proposed in the work of \cite{longo2023adversarial}, while \cite{velazquez2023autonomous} discussed threats to autonomous agents such as MAS from adversarial AI attacks. Similar works to consider adversarial perturbation attacks against maritime radar are \cite{oveis2024advancing} and \cite{du2022practical}. In the optical domain (e.g., digital cameras), the works of \cite{aurdal2019adversarial, lokken2020robustness, pan2024shipcamou} have developed adversarial patches to camouflage ships from single-source AI detection models.

Unlike previous papers examining existing attacks, The work of \cite{walter2024red} used these findings to propose the RedAI framework to support red team evaluations of the cyber security of MAS AI. This is one of the first works to provide a mechanism to help the industry find and mitigate maritime adversarial AI threats. This work provided a test use case to showcase the framework for locating and patching numerous real AAI vulnerabilities in real MAS operating in its true environment. Other security frameworks exist to evaluate the broader state of autonomous cargo ships \cite{yousaf2024cyber}, or audit physical safety \cite{stach2024verifai} and develop safe AI in MAS \cite{yoo2023formulating}.

\subsubsection{Maritime Data Fusion}
Integrating multiple input sources into decision-making processes can yield more robust and potentially more secure models by encompassing a more comprehensive range of information. In marine applications, data is often spatial (e.g., GNSS, sonar, satellite imagery) or temporal (e.g., marine traffic flow) and can be fused using a variety of architectures \cite{munir2021artificial}. Data fusion techniques are classified into low-, intermediate-, and high-level fusion based on the processing stage at which information integration occurs \cite{hall1997introduction}. Low-level data fusion involves combining raw data sources prior to prediction; intermediate-level fusion extracts features from the data for model prediction; high-level fusion entails combining inferences or results from multiple sources to reach a final decision.

Common techniques employed in marine AI data fusion include Bayesian methods \cite{williams2009bayesian, gaglione2020bayesian}, deep learning models \cite{guo2023asynchronous, soldi2021space, duan2022network}, fuzzy logic-based fusion \cite{stover1996fuzzy, liu2021practical}, and Kalman filters or extended Kalman filters \cite{stateczny2011multisensor}. These methods help to overcome uncertainty in noisy, real-world, data. Typical applications involve utilising homogeneous data streams for marine object detection and classification \cite{gulsoylu2024image, bi2024cnngru}, marine environment monitoring \cite{higgins2022ship, xin2024deep}, and marine navigation and tracking \cite{jones2023batman, guo2023asynchronous}. However, most of the current literature on marine AI data fusion focuses on achieving greater precision and reliability for specific marine applications rather than considering data fusion for cyber defence.

Several real-world autonomous ships exemplify the application of these data fusion techniques. Projects such as Rolls-Royce's Advanced Autonomous Waterborne Applications Initiative (AAWA)  \cite{royce2016remote} considers fusing LiDAR, thermal and visual optic data, amongst other sensor data, for AI to enhance autonomous operations. Further, the Mayflower Autonomous Ship is reportedly an AI-powered vessel that uses data fusion from various sensors for transatlantic voyages, sailing the Atlantic autonomously in 2020 \cite{anderson2019bon}. Companies like Robosys are implementing AI and data fusion in maritime systems for autonomous operations. The work of \cite{Barrett2023} developed an AI situational awareness module for remote vessel communication loss. The Yara Birkeland \cite{ziajka2021costs} is the world's first fully autonomous container ship, utilising data fusion for automated coastal hopping. Additionally, the U.S. Department of Defense is also exploring autonomous maritime vehicles to enhance missions.

\section{System Architecture}
\label{sec:modelArch}

\subsection{Data fusion for Situational Awareness}
Across all four IMO degrees of maritime autonomy, there are various applications for AI. Currently degree four, i.e., full autonomy, is defined theoretically, as many legal and technical challenges still have not been overcome. We also note that only some systems need to be fully AI-controlled as this can be a high-risk strategy. In this work, we considered marine AI systems applied to augment a human crew's situational awareness while operating degree three autonomy vessels from a remote operations centre (ROC). 

Real-world AI implementations for situational awareness are more common than other forms of AI for maritime autonomy. Data augmentation and situational awareness are often used to support conventional and remote vessels. There are a plethora of advantages to using these types of AI systems to support remote-controlled vessels where the operator's situational awareness is significantly impaired \cite{thombre2020sensors}. We, therefore, base our initial system on various situational awareness software currently used by real-world vessel operators. This system also allows one to create visual demonstrations to enhance scientific communication. We also highlight the distinction in terminology between the \textit{DFCR method}, which refers to the overarching approach of utilising multiple data sources and fusion techniques to develop defensive AI components, and the term \textit{DFCR system}, which specifically refers to the system evaluated in this paper, created using the DFCR method.

After proving their effectiveness, AI-supported situational awareness may be able to make higher degrees of autonomy more viable in future. For example, this cyber resilient, data fusing system could be used for navigation with a risk model to make the system's decisions more robust and build security into the decision-making process. 

When using AI for high-risk applications (e.g., within CNI, aerial, or marine applications), using a single input source (e.g., optics only) or single modal AI may not be a robust way to operate. The AI model will only use a limited fraction of the information spectrum to make a decision, which may not factor in many important variables (e.g., conditions, environment, security, traffic, political, and social factors), providing a very limited decision. In contrast, a ship's crew use multiple sources of information to make decisions, such as Electronic Chart Display and Information System (ECDIS), visual, radar, Automatic Identification System (AIS), audio cues, and Very High Frequency (VHF) radio. Therefore, AI should also utilise multiple inputs for high-risk decision-making, considering as much relevant information as possible before making a decision. Such information should also be verified where possible to check authenticity and reduce noise.

To integrate multiple data inputs within the DFCR system, we explored data fusion methods that leverage multiple inputs to inform decision-making. These methods enhance system robustness, as single-input models are more vulnerable to being deceived by targeted spoofs or adversarial patches. Since simultaneously spoofing multiple inputs across different sources (e.g., AIS, optics, radar) is significantly more challenging. Nevertheless, the results show that DFCR not only mitigates all single-source attacks but also addresses some of the more complex multi-source attacks.

We test the DFCR system architecture using the RedAI framework  \cite{walter2024red} to assess its vulnerabilities. We discover how multi-pronged attacks can still fool the data-fused AI system, such as a coordinated spoofing of well-positioned AIS messages and a small object's (such as a buoy) radar and optical detections against a basic data fusion system. We, therefore, consider data fusion as a basis for developing more secure systems but build on this work to strengthen the architecture further in the pursuit of creating defence-oriented systems to prevent more sophisticated attacks. 

\subsection{Deriving Defensive Components for AI Systems}

\begin{figure*}[!t]
    \centering
    \includegraphics[width=0.9\linewidth]{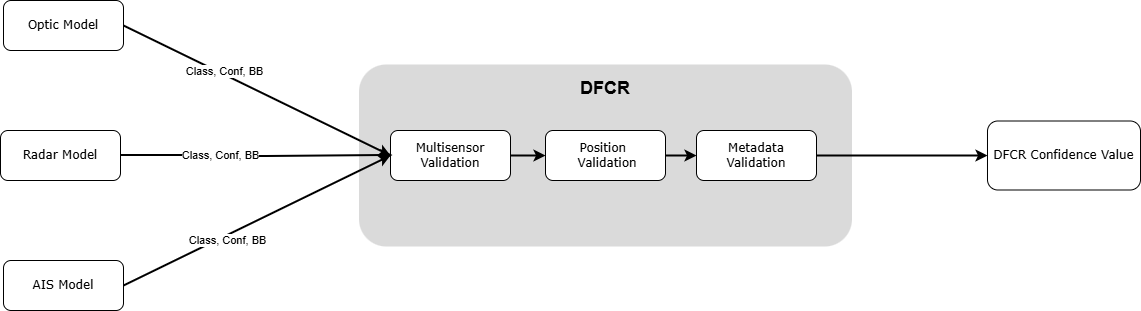}
    \caption{The DFCR system topology shows the defensive components and DFCR confidence output.}
    \label{fig:model_topology}
\end{figure*}

There are many established defensive methods designed to prevent AAI attacks, such as adversarial training \cite{goodfellow2014explaining, madry2017towards} (i.e., a model is trained on adversarial examples) or input preprocessing (e.g., JPEG compression to remove small adversarial perturbations in an image \cite{dziugaite2016study}). 
It is important to note that privacy or model-stealing methods are out of this work's scope as they are less relevant to this particular application.

While many of these defensive methods have been shown to be effective against some attacks, there are often several limitations when facing current adversarial AI methods. Firstly, traditional adversarial AI defences are often restricted to countering a single type of attack. They can lack consistent accuracy, and many only operate effectively within limited or restricted condition (e.g., perturbation size). 

Second, existing metrics that consider security are also limited. Metrics such as model confidence offer limited insight into attacks and are insufficient for integrating security into the system's decision-making process. Existing metrics to measure and understand risk from AAI are very limited, which makes it difficult for developers to improve AAI defences. Furthermore, many current defences are not robust enough to mitigate the effects of attacks at later stages in an AI's system, failing to offer a defence-in-depth approach that is essential for resilient systems. Thirdly, many adversarial AI methods do not address conventional spoofing-based attacks. This oversight leaves systems vulnerable to traditional forms of deception that can compromise system integrity without relying on sophisticated adversarial techniques.

To address these challenges, we have developed a suite of defensive components to create a robust defensive system for the AI system. This approach follows a two-step process:

\begin{enumerate}
    \item Identify potential threats and vulnerabilities: thoroughly analyse the system to identify potential threats specific to AI/ML applications. This includes understanding adversarial attacks, data poisoning, model extraction, and other vulnerabilities unique to AI/ML systems. Utilising a red team framework allows one to simulate attacks and proactively discover weaknesses.  
    
    \item Diversify and enrich system inputs: to mitigate the identified threats, we aim to maximise the diversity and range of data fed into the machine learning system. In the DFCR system, we integrate multiple data sources—including radar, AIS, and optical data, to enhance the system's environmental understanding. This diversity makes it more challenging for an attacker to deceive the system, as they would need to manipulate multiple data types simultaneously. 
\end{enumerate}

For others using this process, step one involves conducting a comprehensive threat assessment to understand the risks pertinent to their specific domains. Identifying these threats to the system enables one to develop defensive components which aim to mitigate these threats. For step two, AI/ML developers should identify and incorporate relevant and diverse data sources pertinent to their specific application areas.

Once these two steps are complete, one can then develop defensive components that utilise the diverse multi-input data to mitigate the identified threats. For example, we can validate and authenticate sensor inputs. To further enhance the system's resilience, we implement robust validation and authentication mechanisms for all input data. This involves verifying the authenticity and consistency of data across multiple sensors and sources. By cross-referencing inputs from radar, AIS, and optical sensors, we can detect anomalies and inconsistencies that may indicate adversarial manipulation and provide a type of anomaly detection. 

For example, consider an attacker attempting a poisoning attack by inserting a backdoor into a model during training and then attaching an optical backdoor trigger resembling an oil tanker to a buoy. With the defensive architecture highlighted in Figure~\ref{fig:model_topology}, this attack is less likely to be successful due to several components:
\begin{itemize}
    \item \textbf{Multisensor validation}: Poisoning a single sensor is not sufficient to fool multisensor situational awareness.
    \item \textbf{Position validation}: The poison trigger would have to be validated against other sensor data, such as positional.
    \item \textbf{Metadata validation}: A trigger designed to mimic an oil tanker would fail if the radar contact does not correspond to that of an actual oil tanker.
\end{itemize}

In another example of data fusion for cyber resilience, a second identified threat might be system accessibility, where defence components focused on redundancy, using diverse inputs, could be implemented to mitigate availability attacks. A third example might include addressing a lack of resilience by enhancing defence-in-depth by utilising the input information and threat assessment to strengthen various layers of the system. By incorporating multiple layers of defence and diverse data sources, the system becomes more robust against attacks that aim to exploit single points of failure. These components could range from simple hard-coded rules to more complex deep neural networks.

Whilst we use a range of models for each of the three sensor inputs, the DFCR system differs from an ensemble approach as it incorporates multiple input data sources, in addition to a data fusion and a security-orientated system backend. We also only use a single model per input source for the initial classification task as well as multiple diverse data sources.  This is unlike an ensemble approach, which would instead run a single data point through multiple object detection models and take a weighted average. 

\subsection{The Experimental Defence Components}
The DFCR system for MAS situational awareness is shown in Figure~\ref{fig:model_topology}. It considers three types of model inputs, each from a different input: AIS, optical, and radar. This data is captured in sequential frames and transmitted to the machine learning system, where a single image displaying the three inputs is generated. This image is then passed through the optical model, the radar model, and the AIS model, which are all object detection models for MAS situational awareness.

For object detection, we utilise YOLOv8 (nano) — open-source models with state-of-the-art benchmark scores \cite{Jocher_YOLO_by_Ultralytics_2023}. We fine-tuned these pre-trained models to recognise AIS, radar, and optical contacts specific to maritime applications. While YOLO models were utilised in this work, the DFCR method is  model-agnostic and can be applied to any set of machine learning models, including Vision Transformers (ViTs) such as DeTR \cite{carion2020end}. YOLO models were selected due to their widespread adoption and prominence as object detection models.

After model inference, each model produces a vector containing information for every detection in each image in the series of images during a voyage. This vector includes class information (Class), confidence scores (C), and bounding boxes (BB) for all contacts. This information is compiled into feature vectors of the form:

\begin{equation} 
\notag
\mathbf{x} = [\text{C}_{m,i}, \ \text{BB}_{m,i}, \ \text{Class}_{m,i}] ,
\label{eq1}
\end{equation}

$\text{For each contact } i \text{ for model } m, \text{where } i \in \mathbb{N}, \text{ and } m \in \mathbb{N} \setminus \{0\} $.  In the case of this work, $m$ is fixed at three sensors and, hence, three models. We then pass these vectors to the defensive components, detailed below, which ultimately recalculate the confidence scores to produce new scores that take into account system security and robustness. 

\begin{figure}[!tbp]
    \centering
    \begin{minipage}[b]{0.45\linewidth}
        \centering
        \includegraphics[height=3cm, trim=100 200 500 50, clip]{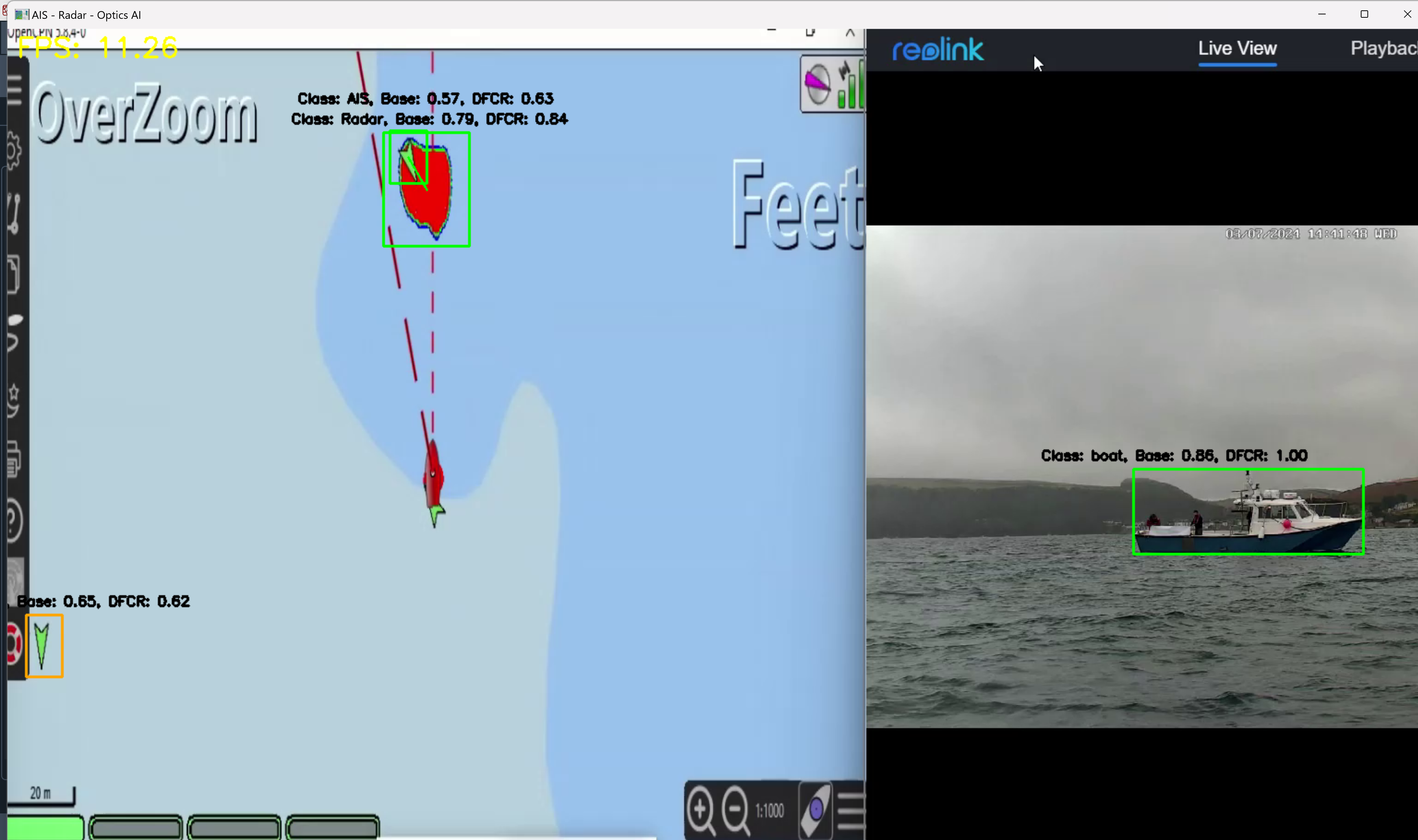} 
        \par\vspace{0.5em} 
        \small (a) A true AIS and radar contact in the AIS and radar coordinate space.
        \label{fig:sec_conf_1}
    \end{minipage}
    \hspace{0.02\linewidth} 
    \begin{minipage}[b]{0.45\linewidth}
        \centering
        \includegraphics[height=3cm, trim=800 100 00 300, clip]{images/system_black.png}
        \par\vspace{1.5em} 
        \small (b) A true optical contact in the optical coordinate space.
        \label{fig:sec_conf_2}
    \end{minipage}
    \caption{The image shows AIS, radar, and optical spaces. A well-verified contact can be seen in both spaces, and this is reflected in improved DFCR confidence scores.} 
    \label{fig:combined}
\end{figure}

In this work, we utilise three defensive components (position validation, multisensor validation, and metadata validation), as detailed in the following sections.  Examples of contacts can be seen in Figures \ref{fig:combined_homography} and \ref{fig:combined}.

\subsubsection{Contact Position Validation and Multisensor Validation}
Once the feature vector $\mathbf{x}$ has been computed by the object detection model, the homography and sector mapping validation component considers the likely positions of detected contacts across different sensor spaces for authentication of contact. If contacts are verified as likely to be the same, for example, radar and corresponding AIS contacts, then their positions within a shared coordinate system should be very close. Contacts may also exist in the optical domain, and a data fusion method known as a homography matrix can be used to map the positions of contacts between these different spaces. The homography matrix can be formally defined as:

A homography matrix \( \mathbf{H} \) is a 3x3 matrix that defines a transformation from one projective plane to another. Given a point \( \mathbf{p} = (x, y, 1)^\top \) in homogeneous coordinates on the first plane, the corresponding point \( \mathbf{p'} = (x', y', 1)^\top \) on the second plane is obtained by: 

\[
\mathbf{p'} = \mathbf{H} \mathbf{p} ,
\]

where:

\[
\mathbf{H} = \begin{bmatrix}
h_{11} & h_{12} & h_{13} \\
h_{21} & h_{22} & h_{23} \\
h_{31} & h_{32} & h_{33} 
\end{bmatrix} .
\]
\vspace{1em}

\begin{figure}[t!]
    \centering
    \begin{minipage}[b]{0.45\linewidth}
        \centering
        \includegraphics[height=3cm, trim=55 130 55 10, clip]{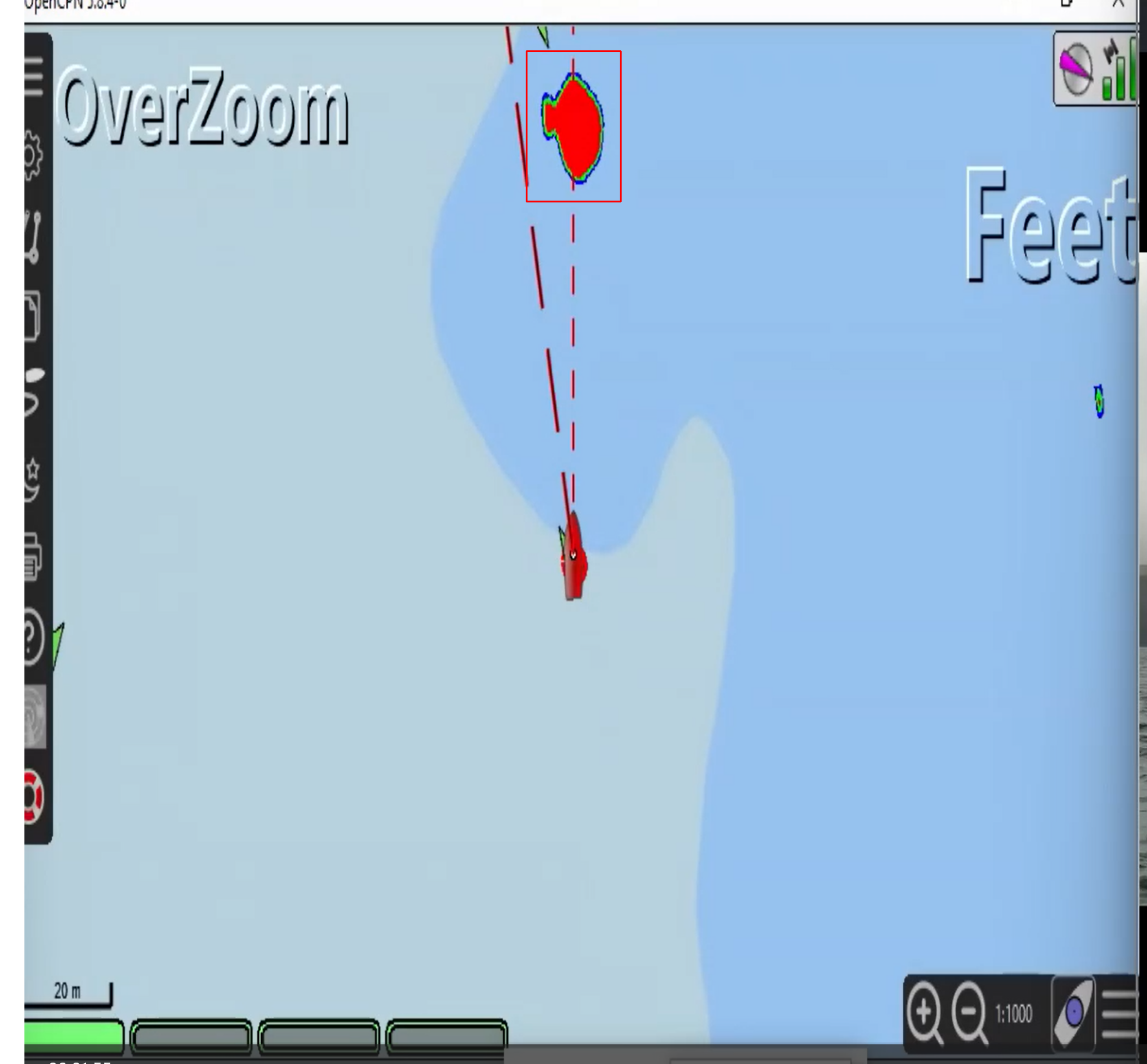} 
        \par\vspace{2em} 
        \small (a) True radar contact (surrounded by a red bounding box) in the AIS and radar coordinate space.
        \label{fig:homography_1}
    \end{minipage}
    \hfill 
    \begin{minipage}[b]{0.45\linewidth}
        \centering
        \includegraphics[height=3cm, trim=50 47 65 95, clip]{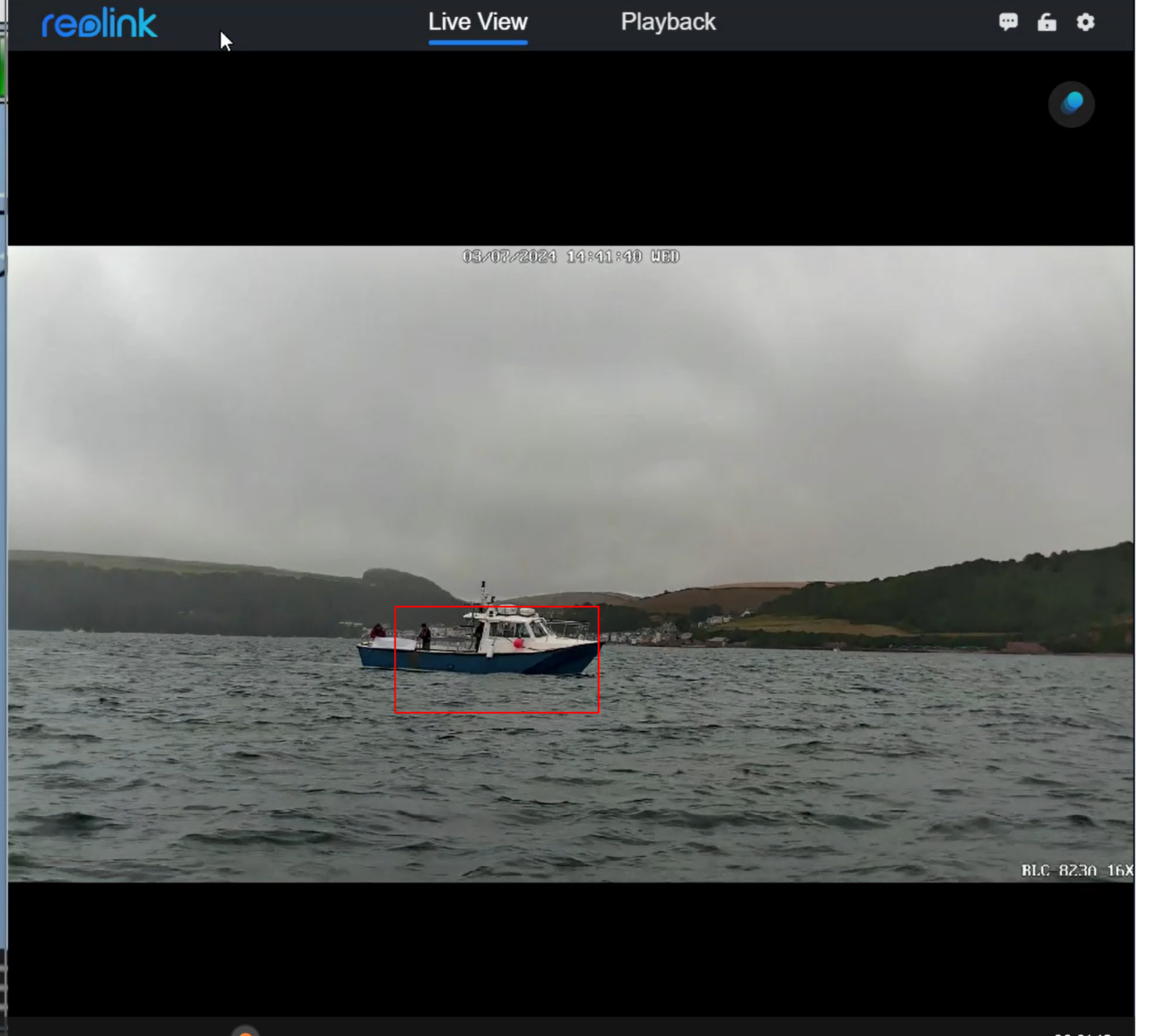}
        \par\vspace{1.0em} 
        \small (b) Radar contact transformed into the optical coordinate space using the homography mapping. Projection is shown as a red bounding box.
        \label{fig:homography_2}
    \end{minipage}
    \caption{Comparison of radar contacts in different coordinate spaces. (a) shows the true radar contact in the AIS and radar space, while (b) shows the radar contact transformed into the optical space using the homography mapping.}
    \label{fig:combined_homography}
\end{figure}

Alternatively, one could develop a more basic coordinate space mapping by splitting the space into sectors and mapping between the two.

Verifying contacts across multiple inputs and ensuring their positions align within a probabilistic expected range can significantly enhance robust decision-making. For example, when an AIS contact and a radar contact are located in close proximity, the DFCR confidence score increases by incorporating this mutual verification. Conversely, if an AIS signal is spoofed, its reported position may not correspond to any radar contact. This scenario would be highly unlikely (if within radar range) unless there is a malfunction of the radar system, the vessel is being spoofed, or it possesses radar return-reducing properties (such as a stealth ship). Such discrepancies would serve as red flags in the decision-making process of the DFCR system.

As seen in Figure~\ref{fig:combined_homography}, radar contacts are transformed using a homography matrix to approximate their positions in the optical space. Given that this method is susceptible to errors, we employ a probabilistic approach to measure confidence levels, utilising a two-dimensional normal distribution centred around each contact. For the DFCR Multisensor Validation component, if contacts that should have corresponding detections (e.g., an AIS report of a ship within radar range) are missing or if contact positions are significantly or unusually misaligned, the system outputs a lower robust confidence score for that object detection. The system first performs multisensor validation by checking for multiple object contacts (e.g., ship) when appropriate and then validates their positions using the contact position.  

\subsubsection{Metadata validation}
Despite utilising multiple inputs to validate each other, it is important to recognise that an attacker could potentially compromise multiple inputs and models simultaneously. For example, an attacker might spoof the AIS signal of an oil tanker and use a strategically placed buoy to create a corresponding radar signature. This scenario highlights the necessity of the metadata validation component in the DFCR system. 

The metadata validation component leverages metadata information, such as a vessel's length and width, from AIS contacts and the signature properties (e.g., size) of the radar contacts, to determine whether these contacts correspond to the same vessel. In the case of a spoofed AIS signal paired with a physical buoy, if the AIS data indicates the contact should be an oil tanker, which is typically a large vessel,  the corresponding radar signature would not match as it would indicate a smaller object. In the DFCR system, such a discrepancy would be flagged as unusual. 

For scenarios such as this, a DFCR component decodes AIS sizing information and compares it with radar size information to assess whether the data may have been compromised. By cross-validating metadata from multiple sensors, the system enhances its ability to identify inconsistencies that may indicate adversarial attacks targeting multiple inputs.

Building upon the critical role of cross-validating metadata from multiple sensors to detect and prevent anomalous activities, we define a system that leverages these multiple inputs for enhanced detection capabilities. During implementation, we considered corresponding contacts from the contact position and multisensor validation check. From this, the matrix $\mathbf{D}$ was produced, where each row contains the corresponding contacts previously matched, and each column contains the relevant metadata features (e.g., contact size). 
This matrix is fed into a Support Vector Machine (SVM). While alternative decision agents such as decision trees, neural networks, random forests, or reinforcement learning algorithms could be used, the SVM provides an effective means of correlating contacts detected by different sensors for this application. 

The objective of the SVM is to determine the probability that each matched contact is either anomalous or plausible by correlating detections across sensor inputs. The SVM classifier can be developed by optimising the loss/objective function to minimise weights $w$ and bias $b$:

\begin{equation*}
\begin{aligned}
&\min _{w, b} \frac{1}{2}\|w\|^2+C \sum_{i=1}^n \max \left(0,1-y_i\left(w \cdot x_i-b\right)\right).
\end{aligned}
\end{equation*}
\vspace{1em}

Here, $C$ is the regularisation parameter that balances the trade-off between correctly classifying each training example and maximising the separation (margin) between classes.
Then, using the optimised $w$ and $b$, the SVM classifier can compute inputs using the decision function:

\[
f(\mathbf{x}) = \operatorname{sgn}(\mathbf{w} \cdot \mathbf{D} + b),
\]

where $\mathbf{w}$ is the weight vector that defines the hyperplane, and $b$ is the bias term, a scalar that offsets the hyperplane. The sign function, $\operatorname{sgn}$, returns +1 if the argument is positive and -1 if the argument is negative, representing the two classes (verified contact or anomalous contact). The classifier produces a prediction of either verification or anomaly, which is subsequently integrated into the final DFCR confidence score. The SVM calculates a decision boundary across a range of features, demonstrating how the SVM differentiates between genuine contacts and potential spoofing attempts.

Robust systems should not only utilise multiple inputs but also leverage information from these inputs to verify the authenticity of the data. By cross-referencing inputs from different sensors, the DFCR approach enhances the system's ability to detect inconsistencies and potential adversarial attacks targeting multiple inputs. This integrated approach improves overall decision-making and resilience against sophisticated threats, as shown in the experimental results (Section~\ref{sec:results}).

\subsection{Secure Metric for AI Defence}
After input data passes through various defensive components, the system calculates and displays the DFCR score passively rather than blocking anomalous contacts. This score, a model confidence metric, integrates security, robustness, situational, and environmental factors into decision-making. For high-risk applications, this information and secure score are relayed to the remote operator to flag unusual behaviour that may require further investigation, helping them or a secondary algorithm make informed decisions. 

The defence component assesses each contact's trustworthiness by outputting a probability or binary result (normal or anomalous), adjusting the confidence score based on a user-defined mapping. For example, if a radar contact aligns with AIS and optical contacts, confidence may increase by 0.3. These adjustments, defined by the developer, consider behaviour probability and unusualness, with certain behaviours penalised more than others. We display this score in Figure~\ref{fig:combined}B. 

As seen in Figure \ref{fig:combined}, the unverified AIS contact confidence is similar to the baseline model confidence. However, the verified radar, AIS, and optical contact for the detected boat have DFCR confidence values that are much higher than the baseline model, reflecting a successful validation through multiple system components. In the visual demonstration, the bounding boxes of multiple authenticated or matched contacts turn green. Further information and visuals could be projected to the operator in future work. We recognise a balance between maximising information and situational awareness without overwhelming the operator \cite{misas2024future}; however, we do not attempt to optimise this in the current work. The DFCR confidence score generation can be seen in pseudocode in Algorithm~\ref{alg:adjusted_confidence} and can be calculated by:

\subsubsection{Initial System Outputs} For each model \( m \) in the set of models \( \{ \text{AIS}, \text{Radar}, \text{Optic} \} \), when an image is passed through the system, we obtain:

\begin{itemize}
    \item \textbf{Confidence Score}: \( C_m^{(0)} \)
    \item \textbf{Bounding Box}: \( \text{BB}_m \)
    \item \textbf{Class Label}: \( \text{Class}_m \)
\end{itemize}

\subsubsection{Validation Components} The initial confidence scores are sequentially provided to three validation components: \\

\begin{adjustwidth}{2em}{0pt}
\noindent\textbf{Component 1: Multisensor Validation}

\noindent\textbf{Objective}: Verify consistency among different models.

\noindent\textbf{Passing Criteria}: Model \( m \) passes if its bounding box \( \text{BB}_m \) and class \( \text{Class}_m \) sufficiently match those from other models. \\

\noindent\textbf{Component 2: Contact Position Validation} 

\noindent\textbf{Objective}: Confirm that the detected contact is within expected positional parameters.

\noindent\textbf{Passing Criteria}: Model \( m \) passes if the contact's position aligns with known or plausible locations. \\

\noindent\textbf{Component 3: Metadata Validation} 

\noindent\textbf{Objective}: Validate additional data associated with the contact.

\noindent\textbf{Passing Criteria}: Model \( m \) passes if the metadata (e.g., vessel size) is correct and consistent. Each component adjusts the confidence score by either penalising or adding a fixed value based on whether the model's output passes the validation.
\end{adjustwidth}
\vspace{1em}

\subsubsection{Confidence Adjustment Mechanism}
\begin{adjustwidth}{2em}{0pt}
\noindent\textbf{Adjustment Amount:} Let \( \delta^{(k)} \) denote the fixed adjustment value for component \( k \), where \( \delta^{(k)} > 0 \).

\noindent\textbf{Passing Indicator:} For each model \( m \) and component \( k \), define the passing indicator:
\end{adjustwidth}

\begin{equation}
\notag
s_m^{(k)} = \begin{cases}
    +1, & \text{if model } m \text{ passes component } k \\
    -1, & \text{if model } m \text{ fails component } k
\end{cases}
\label{eq:passing_indicator}
\end{equation}
\vspace{1em}

\begin{adjustwidth}{2em}{0pt}
\textbf{Confidence Update Equation:} The DFCR confidence score of model \( m \) after passing through component \( k \) is updated as:
\end{adjustwidth}

\begin{equation}
\notag
C_m^{(k)} = C_m^{(k-1)} + \delta^{(k)} \cdot s_m^{(k)}
\label{eq:confidence_update}
\end{equation}

\begin{adjustwidth}{2em}{0pt}
\textbf{Clamping Confidence Scores:} To ensure that confidence scores remain within the valid range \([0, 1]\):
\end{adjustwidth}

\begin{equation}
\notag
C_m^{(k)} = \min\left( \max\left( C_m^{(k)}, \, 0 \right), \, 1 \right)
\label{eq:clamping_confidence}
\end{equation}

\subsubsection{Final DFCR Confidence Score (Combining all updates)}

\begin{equation}
\notag
C_{m}^{\text{final}} = \min \left( \max \left( C_m^{(0)} + \sum_{k=1}^{3} \delta^{(k)} \cdot s_m^{(k)}, \, 0 \right), \, 1 \right)
\label{eq:final_clamping}
\end{equation}

\begin{algorithm}[!t]
\caption{Adjusted DFCR Confidence Calculation for the Defence AI System.}
\label{alg:adjusted_confidence}
\begin{algorithmic}[1]
\REQUIRE Models $M = \{\text{AIS}, \text{Radar}, \text{Optic}\}$ \\
         Initial confidences $C_m^{(0)}$ for each model $m \in M$ \\
         Validation components $K = \{1, 2, 3\}$ \\
         Adjustment amounts $\delta^{(k)} > 0$ for each component $k \in K$ \\
         Passing indicators $s_m^{(k)} \in \{+1, -1\}$ for each model $m$ and component $k$
\ENSURE Final adjusted DFCR confidences $C_m^{\text{final}}$ for each model $m \in M$
\FORALL{models $m \in M$}
    \STATE Initialise confidence: $C_m \leftarrow C_m^{(0)}$
    \FOR{$k = 1$ \TO $3$}
        \STATE Update confidence: $C_m \leftarrow C_m + \delta^{(k)} \times s_m^{(k)}$
        \STATE Clamp confidence: $C_m \leftarrow \min\left( \max\left( C_m, 0 \right), 1 \right)$
    \ENDFOR
    \STATE Store final adjusted DFCR confidence: $C_m^{\text{final}} \leftarrow C_m$
\ENDFOR
\RETURN $C_m^{\text{final}}$ for each model $m \in M$
\end{algorithmic}
\end{algorithm}

\section{Experimental Setup}
\label{sec:method}
One of the objectives of this work is to develop defensive components that contribute to the generation of a DFCR score. These components and the DFCR score were evaluated through two distinct methodologies. Firstly, a series of real-world demonstrations were conducted to assess the practical impacts of attacks and defences on these systems. The practical findings and limitations fed into the analysis provided in the experimental section and discussion. Secondly, a set of controlled experiments were performed to quantitatively evaluate the defensive systems as practical defence methods. These experiments compare the DFCR approach against single-input models and models utilising existing state-of-the-art defences.

The defences selected for this study are the most established and commonly used methods, tailored to be applicable to specific attack types. For instance, JPEG compression defences are not considered for defending against AIS spoofing attacks, as such an approach lacks logical applicability. The chosen defences are relevant to each targeted attack. These defences include compression and input preprocessing (e.g., JPEG compression) and adversarial training applied to the single-input models. This selection allows for a comparison and benchmarking of the defensive system's effectiveness against some of the most popular current state-of-the-art defences.

The DFCR system and models were tested against a range of the most prevalent and pertinent attacks identified in the background literature. Privacy-based attacks are excluded from this study as they fall outside the scope of the model, data, and application context. We utilised the RedAI framework \cite{walter2024red} to identify AI vulnerabilities and attacks that could be used to evaluate the DFCR system. 

From RedAI, the attacks considered to test the situational awareness AI include adversarial patches, adversarial perturbations, and sensor spoofing (AIS and radar jamming/reflection/electronic warfare simulations). These attack types will constitute four separate experiments intended to assess the DFCR system developed for MAS situational awareness. During these attacks, the confidence values of different systems and models, including the DFCR confidence score, will be compared to measure the effectiveness of the defences.

\subsection{Marine Dataset and Equipment}
\begin{figure}
    \centering
    \includegraphics[width=1.0\linewidth]{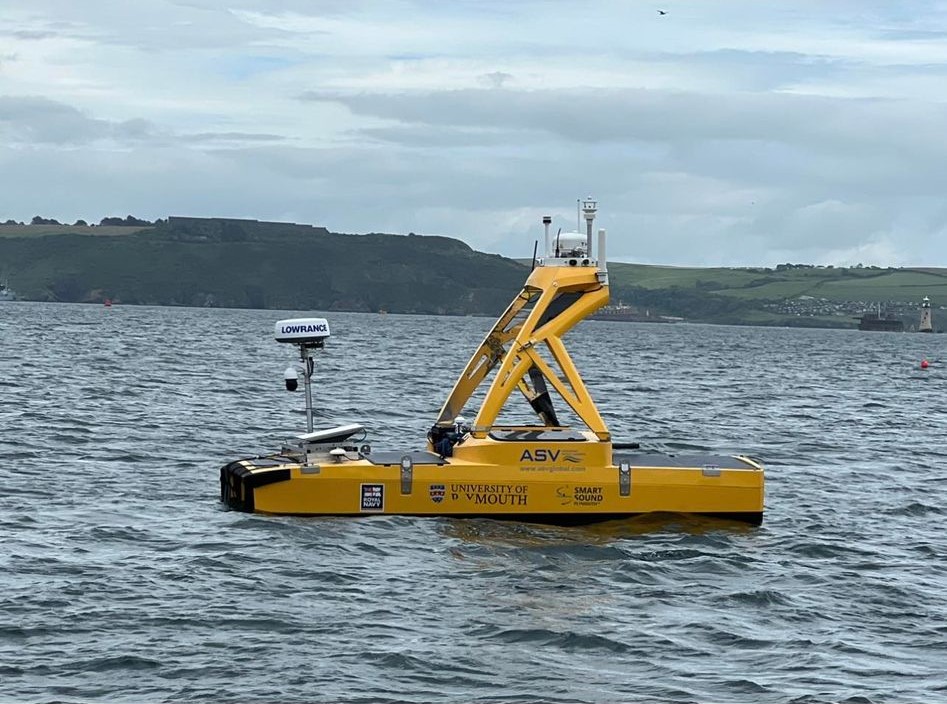}
    \caption{The USV Bauza.}
    \label{fig:Bauza}
\end{figure}

All data utilised in this study was collected using the Uncrewed Surface Vessel (USV) Bauza (C-Enduro), an autonomous experiment platform operated by the University of Plymouth. Typically managed remotely from a ROC, the vessel's operation inherently limits the operator's situational awareness. The system developed in this work aims to enhance crew situational awareness by leveraging and combining the vessel's sensor capabilities. USV Bauza (see Figure~\ref{fig:Bauza}) serves dual purposes: it is the source of training and validation data and the platform for evaluating model inference and conducting real-world AI defence simulations. 

Data collection was conducted in the Cawsand USV range at Plymouth Smart Sound, a distinctive body of water within UK territory that facilitates the safe deployment of marine autonomous equipment. Data acquisition spanned multiple days (2022-2024) and encompassed a variety of scenarios to ensure a comprehensive and diverse dataset. The dataset comprises screen recordings of radar, 4K optical (camera), navigational charts, and AIS data. All data was manually labelled, with the detection confidence initially set to the default YOLO value of 0.3. Most experimental parameters remained at their default settings unless adjustments were necessary; any modifications and their justifications are detailed in the experimental section. For real-world application of the defences and models, considerations regarding risk appetite and specific use cases should guide parameter settings.

\section{Experimental Results}
\label{sec:results}

Four experiments were conducted to demonstrate the effectiveness of the DFCR method. The attacks used to test defences were derived from the RedAI framework to find the most appropriate AAI attacks for evaluation. The experiments were conducted using the following hardware configurations: 
\begin{itemize}
    \item Primary Inference System: Intel Core i9-13900H CPU, 16 GB DDR5 RAM, and NVIDIA RTX 4070 GPU. 
    \item Development Environment: Google Colab with an Intel Xeon CPU at 2.20 GHz, NVIDIA A100-SXM4-40GB GPU, and 51 GB system RAM.
\end{itemize}

A key terminology clarification for the upcoming sections is that \textbf{DFCR system confidence} refers to the confidence output from the DFCR-enhanced system. In contrast, \textbf{ baseline model confidence} refers to the confidence derived from the standalone models (i.e., the same object detection model but without the DFCR defensive components).

\subsection{Experiment 1: Clean Performance}
This experiment measured differences (improvement or depreciation) between the baseline model confidence score and the newly proposed DFCR confidence score in normal operating conditions (whilst not being attacked). We selected 300 distinct scenarios, represented by screenshots (images) of the optical and navigational interfaces, and processed each scenario through both the DFCR system and the baseline model. The resulting DFCR system confidence and baseline model confidence scores were recorded for analysis and comparison.

We used a range of metrics centred around loss. Loss is the difference between what is true (e.g., there is a real boat contact in range of the MAS situational awareness AI) and what has been predicted by the system (i.e., predicting a high confidence of boat contact). Therefore, a lower loss value is more desirable, as the system prediction would be as similar as possible to the truth. This is different from raw values (raw confidence), which will depend on whether or not a true contact exists. For example, if a contact is spoofed, a better system should produce a lower contact score for that spoof while producing high confidence values for the true contacts. 

The confidence scores of the baseline model and the DFCR system for each metric are presented in Table~\ref{tab:metrics_comparison_2}. The loss values are either nearly identical to or lower for the DFCR confidence score, indicating that the DFCR system performs better. Each scenario includes a number of correct contacts; therefore, a lower loss score signifies that the system is more effective at providing confidence for true contacts. In this work, we display both Mean Squared Error (MSE) and Mean Absolute Error (MAE). MSE penalises larger errors more severely, whilst MAE penalises errors in a more linear way. However, another developer may choose to pay particular attention to one metric or the other depending on the risk/attention to larger errors. 

\begin{figure}[!tbp]
    \centering
    \begin{tikzpicture}
        \node at (-4,0) {\includegraphics[width=0.9\linewidth]{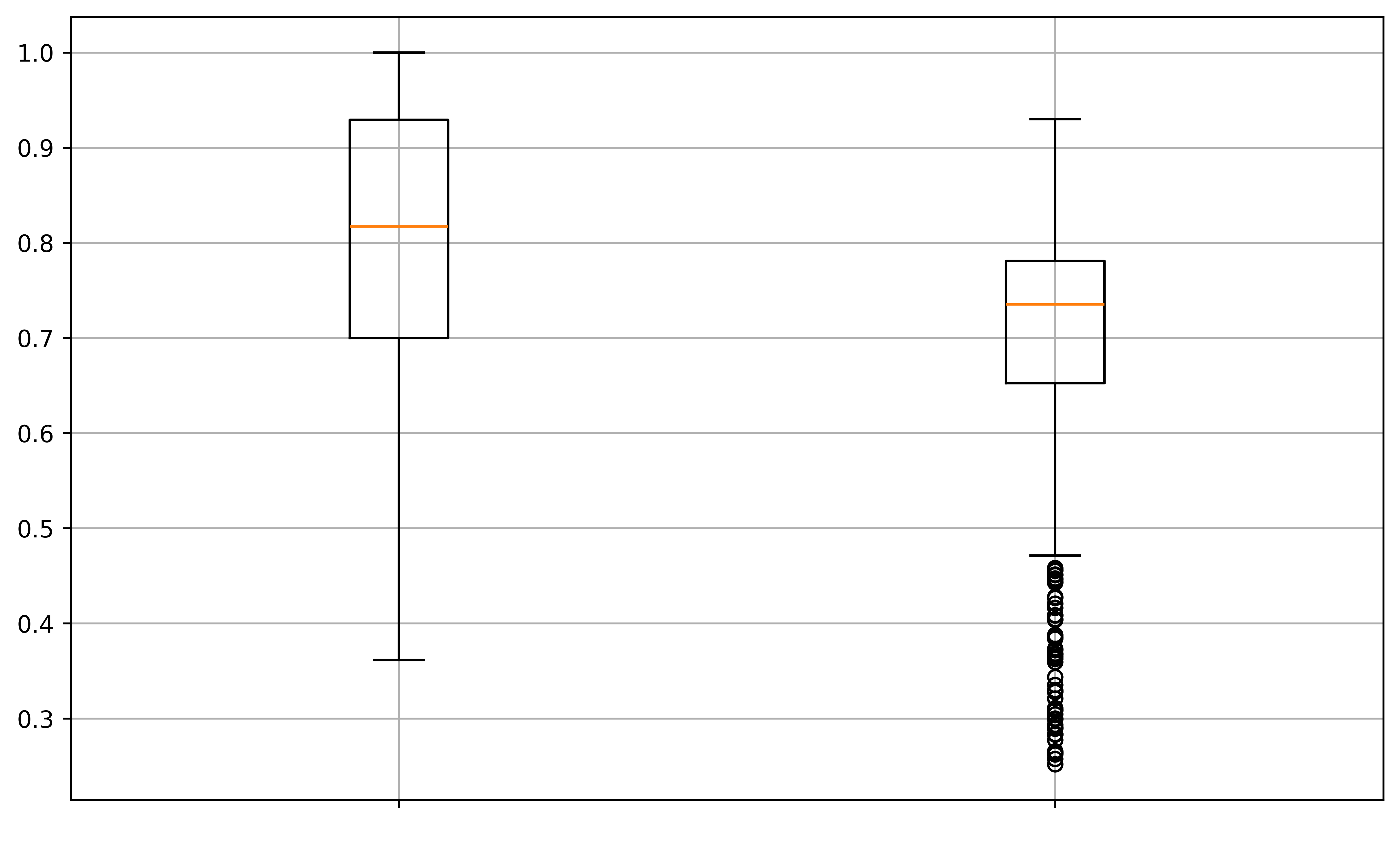}};
        
        \node[rotate=90, anchor=center] at (-8.2,0) {\bf\footnotesize Raw Confidence Values};
        
        \node[anchor=center] at (-5.5,-2.5) {\bf\footnotesize DFCR Confidence};
        \node[anchor=center] at (-2.2,-2.5) {\bf\footnotesize Baseline Confidence};
    \end{tikzpicture}
    \caption{Elevated $y-$values (raw confidence values) correspond to superior detection capabilities, as all detections are genuine.}
    \label{fig:box_plot_e1}
\end{figure}
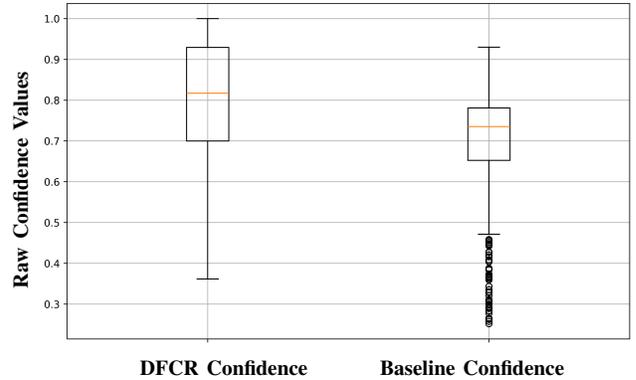

\begin{table}[!tb]
\centering
\caption{Comparison of metrics between DFCR confidence and baseline model confidence. (Lower Values Are Better) under normal conditions.}
\begin{tabular}{p{3.8cm}p{1.6cm}p{1.6cm}}
\toprule
\textbf{Metric} & \textbf{DFCR Conf} & \textbf{Baseline Conf} \\ \midrule
MSE Loss & 0.1211 & 0.1713 \\ 
RMSE Loss & 0.3480 & 0.4139 \\ 
Median of Differences & 0.2195 & 0.3035 \\ 
Range of Differences & 0.7747 & 0.6188 \\ 
Std Dev of Differences & 0.2421 & 0.1692 \\ 
MAE & 0.2500 & 0.3777 \\ 
\bottomrule
\end{tabular}
\label{tab:metrics_comparison_2}
\end{table}

As seen in the initial test, both the DFCR system and baseline models exhibit low loss values, indicating that the baseline model already performs well in these conditions. However, the DFCR system's confidence achieves a 30\% reduction in MSE loss (0.12) compared to the baseline model (0.17) due to the increased availability of information, such as a higher number of contacts and multiple input modes. This abundance of data allows the system to utilise its defensive components. Additionally, the MAE and Root Mean Squared Error (RMSE) are approximately one-quarter lower for the DFCR system's confidence, underscoring a significant improvement in the detection and verification of contacts. In Figure~\ref{fig:box_plot_e1}, the box plot shows the improved $y-axis$ scores for the DFCR system's confidence. 

Given the presence of outliers and the non-normal distribution of the data, we employed the Wilcoxon signed-rank test, a non-parametric method that uses ranks to assess the median differences between two related groups. This approach produced a $p$-value of  \(7.291 \times 10^{-75}\), far below the conventional significance threshold of 0.05. These results strongly suggest that the observed improvements are not due to random chance, confirming the statistically significant differences between the two methods.

In situations of low-activity scenarios, we anticipate that the DFCR confidence and traditional confidence scores will be more similar, as situations with few contacts and verifications do not fully capitalise on the DFCR confidence components, such as verification, resulting in the system operating at reduced defence effectiveness. The analysis from these tests demonstrates that the DFCR system generally outperforms the baseline model across various metrics. Specifically, the DFCR system's confidence exhibits lower errors in MSE, RMSE and MAE. These findings underscore the effectiveness of the DFCR method in improving legitimate detection and verification capabilities in normal-activity environments, thereby providing a more reliable and robust system.

\subsection{Experiment 2: Perturbation Attack Defence}
Building upon the benchmark comparison between the baseline model and the DFCR system using benign data, which demonstrated the DFCR system's robustness under normal operating conditions, experiment two evaluated the DFCR system's performance under adversarial AI attack scenarios. We generated adversarial perturbations on the input image, which would fool the system into detecting objects (e.g., a radar contact) that do not really exist. The objective of the attacker may be to fool the AI vessel into detecting objects that do not exist in real space and, hence, confuse or change the trajectory of the vessel. We then tested the DFCR system to see if it could flag adversarial perturbations added to inputs by providing a very low or zero confidence score to the operator.

Various methods exist for generating adversarial perturbations. Open-box methods, such as the Fast Gradient Sign Method (FGSM) and Projected Gradient Descent (PGD), require access to the model's gradients \cite{goodfellow2014explaining, madry2017towards}. Conversely, we employ a black-box or closed-box approach using an evolutionary algorithm (EA), specifically NSGA-III \cite{deb2013evolutionary}, to generate adversarial perturbations without necessitating gradient calculations \cite{williams2024evolutionary}. 

\begin{figure}[!tbp]
    \centering
    \begin{tikzpicture}
        \node at (-4,0) {\includegraphics[width=1.0\linewidth]{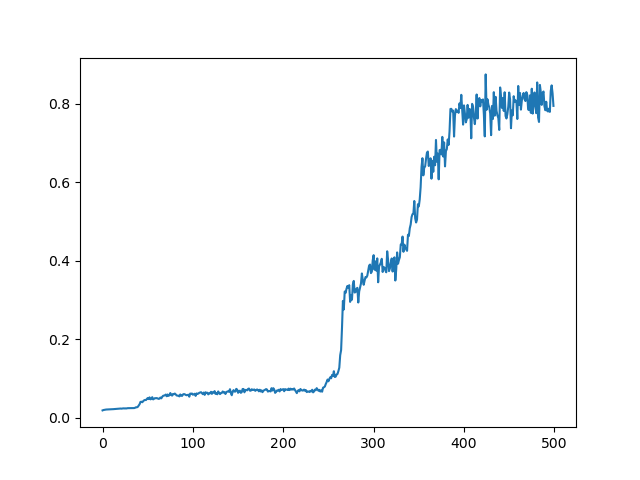}};
        
        \node[rotate=90, anchor=center] at (-8,0) {\bf\footnotesize Average Fitness Score };
        
        \node[anchor=center] at (-3.8,-3.2) {\bf\footnotesize Iterations};
    \end{tikzpicture}
    \caption{An evolutionary algorithm (EA) evolving adversarial patches for perturbation attacks, illustrating the average fitness score of 50 individuals over 500 iterations.}
    \label{fig:fitness}
\end{figure}

The perturbation generation process utilises image pixels as the parameter space and the model confidence scores for AIS and radar as the objective (fitness) functions, with the goal of maximising these confidence scores. NSGA-III was selected for its robust capability to identify the Pareto front in many-objective optimisation problems, allowing for the expansion of objectives to include more nuanced criteria if needed. Figure~\ref{fig:fitness} illustrates an example of the EA evolving solutions to maximise the combined model confidence. Table~\ref{tab:parameter_settings} outlines the hyper-parameter settings of the EA used in this study to facilitate reproducibility.

\begin{table}[!tbp]
\centering
\caption{Hyper-parameter settings for the optimisation algorithm.}
\begin{tabular}{p{4cm}p{1cm}}
\toprule
\textbf{Hyper-Parameter} & \textbf{Value} \\ \midrule
Number of Iterations (Max) & 500 \\ 
Number of Iterations (Min) & 50 \\ 
Population Size & 50 \\ 
Perturbation Size ($\epsilon$) & 50 \\ 
Decay Factor for Epsilon & 0.9 \\ 
No Improvement Threshold & 30 \\ 
 
\bottomrule
\end{tabular}
\label{tab:parameter_settings}
\end{table}

\begin{table*}[!t]
\centering
\caption{Comparison of Metrics between Different Systems and Model Confidence (Lower Values Are Better) during an adversarial perturbation attack.}
\begin{tabular}{p{2.8cm}p{1.8cm}p{1.8cm}p{1.8cm}p{1.8cm}}
\toprule
\textbf{Metric} & \textbf{Baseline Confidence} & \textbf{Secure Confidence} & \textbf{JPEG Confidence} & \textbf{Noise Confidence} \\ \midrule
MSE Loss & 0.4994 & 0.3231 & 0.1951 & 0.4985 \\ 
RMSE Loss & 0.7066 & 0.5684 & 0.4417 & 0.7061 \\ 
Median Difference & 0.7591 & 0.4554 & 0.3952 & 0.7587 \\ 
Range of Differences & 0.7746 & 0.9229 & 0.6729 & 0.7812 \\ 
Std Dev of Differences & 0.1326 & 0.1908 & 0.2985 & 0.1335 \\ 
MAE & 0.6941 & 0.5354 & 0.3256 & 0.6933 \\ 
\bottomrule
\end{tabular}
\label{tab:metrics_comparison_confidence_models_ex3}
\end{table*}

The perturbations generation can be formulated as a multi-objective optimisation problem. More formally, 
\[
\begin{aligned}
&\text{Maximise} \quad \mathbf{F}(\mathbf{x}) = \left( f_1(\mathbf{x}), f_2(\mathbf{x}), \dots, f_M(\mathbf{x}) \right) \\
&\text{Subject to} \quad \mathbf{x} \in \Omega
\end{aligned}
\]
where \(\mathbf{x} = \left( x_1, x_2, \dots, x_n \right)\) is the decision vector, \(\mathbf{F}(\mathbf{x})\) is the objective vector consisting of \( M \) objective functions (for this work $M=2$, and \(\Omega\) is the feasible decision space defined by constraints. Before each solution in the population is evaluated, a 0-255 clip is applied to ensure the perturbation values remain within an appropriate range. The number of generations is randomly selected for each situation between 50 and 500 generations.

\begin{figure}[t!]
    \centering
    \begin{tikzpicture}
        \node at (-6,0) {\includegraphics[width=1\linewidth]
        {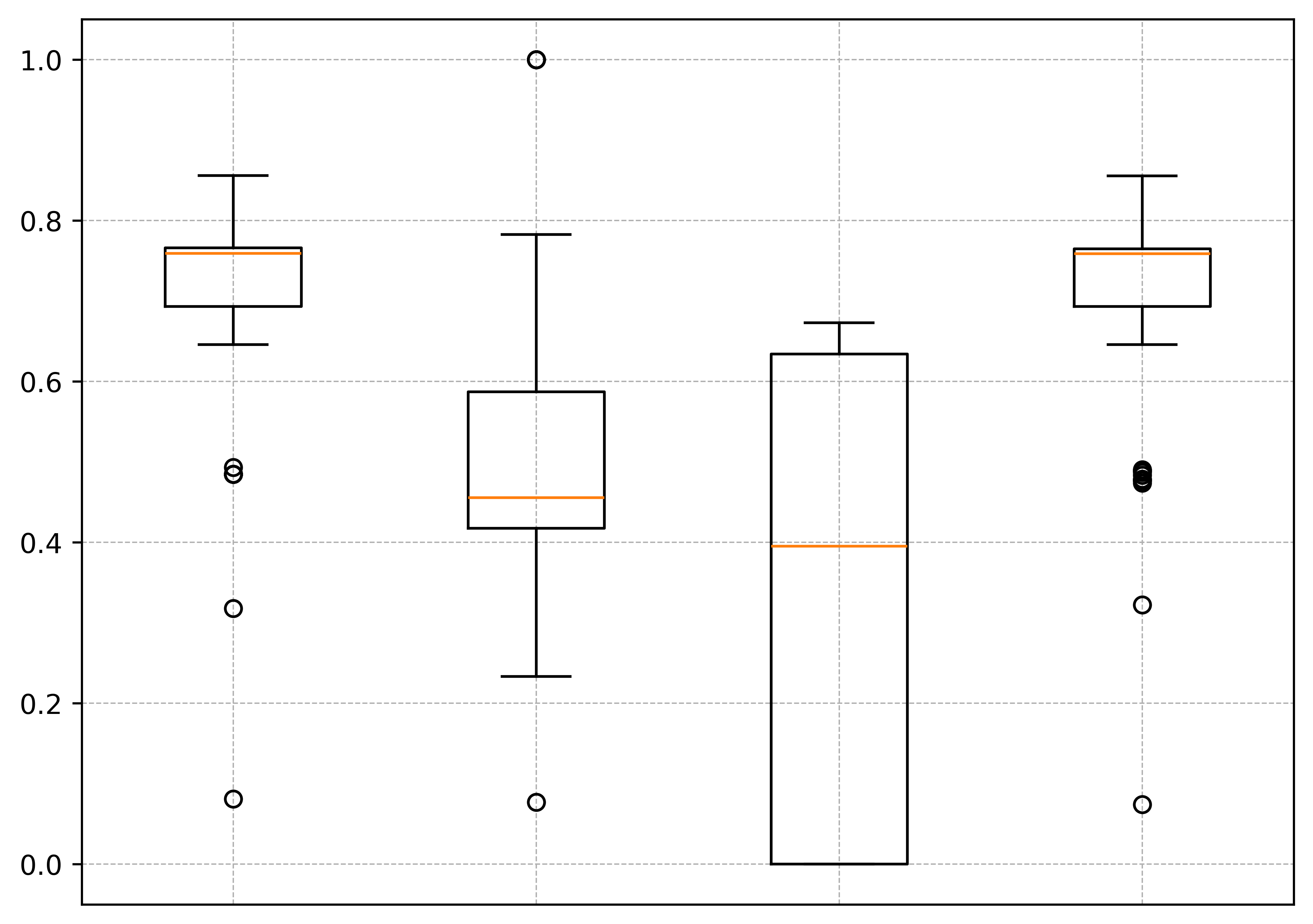}};
        
        \node[rotate=90, anchor=center] at (-10.7,0) {\bf\footnotesize Raw Confidence Values};
        
        \node[anchor=center] at (-4.8,-3.2) {\bf\footnotesize JPEG};
        \node[anchor=center] at (-8.9,-3.2) {\bf\footnotesize Baseline};
        \node[anchor=center] at (-6.85,-3.2) {\bf\footnotesize DFCR};
        \node[anchor=center] at (-2.7,-3.2) {\bf\footnotesize Noise};

        \node[anchor=center] at (-4.8,-3.5) {\bf\footnotesize Conf};
        \node[anchor=center] at (-8.9,-3.5) {\bf\footnotesize Conf};
        \node[anchor=center] at (-6.85,-3.5) {\bf\footnotesize Conf};
        \node[anchor=center] at (-2.7,-3.5) {\bf\footnotesize Conf};
    \end{tikzpicture}
     \caption{A box plot illustrating the preliminary confidence scores of systems and models subjected to various defence mechanisms. Lower $y$-values indicate reduced confidence means, with a score of 0 representing optimal performance during adversarial perturbation attacks.}
    \label{fig:exp3_boxplot}
\end{figure}
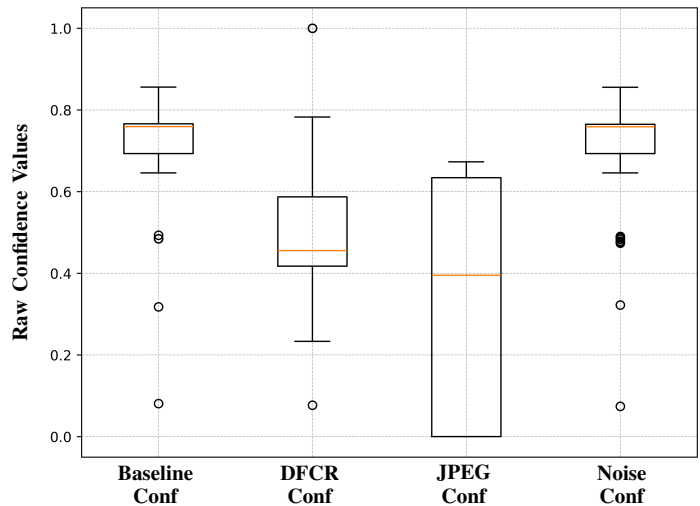

This evaluation involves comparing the DFCR system's confidence against the baseline model confidence, as well as against a set of baseline models that incorporate state-of-the-art adversarial defences from the literature. These defences include sterilisation and compression methods seek to eliminate adversarial perturbations from inputs by reducing their resolution, thereby potentially removing noise and distortions.

The DFCR system's confidence, the confidence of the baseline model and the confidence of the baseline model with defences are provided in Figure~\ref{fig:exp3_boxplot} and Table~\ref{tab:metrics_comparison_confidence_models_ex3}. In this context, lower loss values indicate better performance, as they reflect reduced or negligible confidence in adversarial attacks, thereby rendering the attacks unsuccessful or diminishing their impact on the system. The defences implemented in this study include JPEG compression and Gaussian noise addition techniques. In these experiments, we utilised a sample size of 100 scenarios. 

As illustrated in Figure~\ref{fig:exp3_boxplot} and Table~\ref{tab:metrics_comparison_confidence_models_ex3}, the DFCR confidence was one of the better defences in all metrics. Specifically, the JPEG compression defence was the only alternative that effectively reduced attack perturbations; however, its efficacy was limited to perturbation-based attacks alone. The raw confidence standard deviation of the JPEG compression defence is illustrated in Figure~\ref{fig:exp3_boxplot} and can be seen to be around three times higher than that of other systems and hence less consistent. Other compression algorithms demonstrated reduced effectiveness, likely due to the perturbations being too large, allowing their effects to persist even after defence application. While it is theoretically possible to increase the level of compression to eliminate larger perturbations, such an approach would likely compromise the quality of the original images, thereby negatively impacting the detection of legitimate contacts. In contrast, the method we propose offers the advantage of maintaining original image quality, thereby preserving the accuracy of benign detections.

In marine detection applications, objects at a distance typically appear small due to the camera's focal length and the challenges inherent in operating within expansive, open-water environments. Compression algorithms, particularly those designed to reduce image size and bandwidth, often achieve this by minimising less noticeable details, which can include small objects in the background. Consequently, essential detections, such as distant vessels or buoys, may be compressed into the background and go undetected. This issue poses a significantly larger problem than adversarial attacks, as it directly affects the system's core functionality and reliability. Furthermore, work in \cite{shin2017jpeg} developed JPEG-resistant adversarial images, limiting the impact of the JPEG-compression defence.

The Wilcoxon test yielded a p-value of 3.412 \(\times 10^{-8}\). Similar to Experiment 1, the Wilcoxon test value suggests that the DFCR system's median confidence differs from the baseline's median confidence. These results indicate that the observed differences between the DFCR system and baseline model confidence are statistically significant at the conventional alpha level of 0.05. Overall, the DFCR method shows meaningful improvements in performance metrics. Furthermore, the DFCR system exhibits lower errors in MSE, RMSE, and MAE than the baseline model, along with comparable or lower values for other performance metrics. 

\subsection{Experiment 3: Patch Attack Defence}
We now consider the system’s robustness against adversarial patch attacks. An attacker could use a digital or physical adversarial patch to manipulate the vessel's behaviour, potentially causing it to change its trajectory or take an unusual action. The attacks for this experiment were generated with the Projected Gradient Descent (PGD) method \cite{madry2017towards}. We assume an open-box adversarial setting where the attacker has access to the model's gradient to carry out the PGD attack. The PGD attack can be formulated such that the adversarial example \( \mathbf{x}_{\text{adv}} \) is crafted as:
\[
\mathbf{x}_{\text{adv}} = \mathbf{x} + \epsilon\, \operatorname{sgn}\left( \nabla_{\mathbf{x}} J(\theta,\, \mathbf{x},\, y) \right) .
\]
Here, \( \mathbf{x} \) represents the original image, \( \epsilon \) is the perturbation size, and \( \operatorname{sgn}\left( \nabla_{\mathbf{x}} J(\theta, \mathbf{x}, y)\right) \) indicates the direction of the gradient aimed at maximising the model's confidence for the given input so that the model detects non-existent detections. The objective of the PGD attack is to maximise the model's confidence (or equivalently minimise the loss) as follows:
\[
\begin{aligned}
&\text{Minimise} \quad \mathbf{F}(\delta) = \left( -J(\theta,\, \mathbf{x} + \delta,\, y),\ \|\delta\|_p \right) , \\
&\text{Subject to} \quad \mathbf{x} + \delta \in \Omega ,
\end{aligned}
\]
where:
\begin{itemize}
    \item \( \delta \) is the perturbation.
    \item \( -J(\theta,\, \mathbf{x} + \delta,\, y) \) aims to maximise the loss.
    \item \( \|\delta\|_p \) measures the magnitude of the perturbation.
    \item \( \Omega \) ensures inputs remain within a valid domain.
\end{itemize}
\vspace{1em}

This experiment focuses on attacking only the optical detection model with adversarial patches generated using PGD. We employ PGD parameters with \( \alpha = 0.05 \), ten iterations, and \( \epsilon = 0.3 \). To defend against such attacks, we use adversarial training, where an additional 197 adversarial patches are generated and incorporated into the training dataset of the optical detection model. This defence method improves the model's robustness by introducing training data representative of adversarial examples. Specifically, about 10\% of the training dataset consists of adversarial data. The model was retrained for 100 epochs with a batch size of eight, enhancing its ability to withstand adversarial patch attacks.

\begin{table}[!t]
\centering
\caption{Comparison of metrics between the baseline model confidence, the DFCR system confidence, and the adversarial trained model confidence during an adversarial patch attack.}
\begin{tabular}{p{2.8cm}p{1.2cm}p{1.2cm}p{1.2cm}}
\toprule
\textbf{Metric} & \textbf{Baseline} & \textbf{DFCR} & \textbf{Adversarial Trained} \\ \midrule
MSE Loss & 0.2542 & 0.0000 & 0.1990 \\ 
RMSE Loss & 0.5042 & 0.0000 & 0.4461 \\ 
Median of Differences & 0.4910 & 0.0000 & 0.4380 \\ 
Range of Differences & 0.3741 & 0.0000 & 0.6425 \\ 
Std Dev of Differences & 0.0699 & 0.0000 & 0.1141 \\ 
MAE & 0.4993 & 0.0000 & 0.4313 \\ 
\bottomrule
\end{tabular}
\label{tab:metrics_comparison_conf_secure_adv}
\end{table}

Table~\ref{tab:metrics_comparison_conf_secure_adv} summarises the results of experiment three. The three table columns represent the baseline model confidence, the DFCR confidence, and the adversarially trained model confidence. The DFCR confidence exhibits the smallest squared error (0.00) by a significant margin, indicating that the system's loss during an adversarial optical patch attack was the lowest. This suggests that the DFCR system effectively disregarded or was robust to, these adversarial attacks. The adversarial model demonstrated the second most effective performance but only showed an improvement of approximately 0.06 in MSE loss compared to the baseline confidence model. These trends are consistent across other metrics and raw values.

Furthermore, the statistical Wilcoxon test to compare median differences between the baseline model and the DFCR system yield p-values of  \(8.329 \times 10^{-18}\), which confirms that the observed differences are statistically significant and not due to random chance.

Furthermore, while the adversarially trained defence did lead to a slight reduction in the normal accuracy of the model, the DFCR method did not alter the original model performance, unlike adversarial training, which usually requires a trade-off to improve robustness to adversarial attacks at the cost of lower model accuracy on true detections. Hence, the initial model detection accuracy did not diminish in the DFCR system. Much of the DFCR system's defence likely relies on the absence of corresponding AIS or radar inputs to validate contacts identified by the optical model. Consequently, contacts without radar verification, despite being large enough and within the radar's range, were effectively disregarded by the DFCR system.

In summary, the analysis shows that the DFCR system provides the best performance. The DFCR system's confidence metrics all returned zero, effectively disregarding these adversarial inputs. The adversarially trained model does improve upon the baseline model confidence by reducing certain types of errors but introduces greater variability in others. Overall, the DFCR system achieved the highest accuracy and the least error across the tested metrics.

\subsection{Experiment 4: AIS and Radar Spoof Defence}
In this final experiment, we evaluate the system's resilience to  AIS and radar spoofing. AIS spoofing involves injecting false AIS signals directly into the MAS system, while radar spoofing entails adding deceptive radar contact signals to the scenarios/images. Details on AIS spoofing in the marine domain can be found in \cite{kessler2024ais}. The unsecured nature of the AIS protocol, based on NMEA protocols, makes AIS spoofing one of the most straightforward attacks to develop and test. Defending against AIS and radar spoofing is particularly challenging, as conventional defences such as compression or adversarial training are ineffective against these types of attacks. Therefore, we focus solely on evaluating the system's intrinsic defensive components without comparing them against external defence mechanisms. Each spoofed AIS or radar signal that does not match the correct probabilistic signature results in a lower loss score, as the metadata validation process should penalise detections with significant mismatches. 

This experiment was designed to introduce a range of radar and AIS spoofed signals per scenario to maximise detection potential and enable the system's defensive components to perform verification checks. The total number of AIS and radar detections per scenario is limited to one, three and five. Each test comprises 100 examples for each number of spoofed signals to ensure statistical robustness. 

\begin{table*}[!tbp]
\centering
\caption{Comparison of performance metrics between the DFCR system's confidence and the baseline model confidence under conditions with 1, 3, and 5 AIS/radar spoofed signals. Lower values indicate improved performance, signifying reduced confidence in adversarial attacks and enhanced model/system robustness.}
\renewcommand{\arraystretch}{1.3}  
\begin{tabularx}{\textwidth}{p{3.5cm} p{1.9cm} p{1.9cm} p{1.9cm} p{1.9cm} p{1.9cm} p{1.9cm}}
\toprule
\textbf{Metric} & \multicolumn{2}{c}{\textbf{1 Combination}} & \multicolumn{2}{c}{\textbf{3 Combinations}} & \multicolumn{2}{c}{\textbf{5 Combinations}} \\ 
\cmidrule(lr){2-3} \cmidrule(lr){4-5} \cmidrule(lr){6-7}
& \textbf{DFCR Confidence} & \textbf{Baseline Confidence} & \textbf{DFCR Confidence} & \textbf{Baseline Confidence} & \textbf{DFCR Confidence} & \textbf{Baseline Confidence} \\ \midrule
\textbf{MSE Loss} & 0.0000 & 0.5128 & 0.5136 & 0.6446 & 0.4441 & 0.6217 \\ 
\textbf{RMSE Loss} & 0.0000 & 0.7161 & 0.7167 & 0.8028 & 0.6664 & 0.7885 \\ 
\textbf{Median of Differences} & 0.0000 & 0.7408 & 0.7745 & 0.8359 & 0.5581 & 0.8310 \\ 
\textbf{Range of Differences} & 0.0000 & 0.7069 & 0.8086 & 0.7427 & 0.9146 & 0.8327 \\ 
\textbf{Std Dev of Differences} & 0.0000 & 0.1198 & 0.1777 & 0.1043 & 0.1839 & 0.1272 \\ 
\textbf{MAE} & 0.0000 & 0.7060 & 0.6943 & 0.7960 & 0.6406 & 0.7781 \\ 
\bottomrule
\end{tabularx}
\label{tab:metrics_combined_comparison}
\end{table*}

As presented in Table~\ref{tab:metrics_combined_comparison}, when attacked by a single spoofed contact, the MSE of the DFCR system's confidence (0.00) is significantly better than the baseline model confidence (0.51). This is likely due to missing but expected corresponding contacts that can validate the spoofed contact. This indicates that the DFCR method has significantly reduced the impact of spoof attacks attempting to fool the AI system. 

We can observe that as the number of spoofed contacts increases, the DFCR system receives more information for decision-making, such as additional verification data, allowing the defensive components to operate more effectively, improving the system's performance metrics (enhancing defence effectiveness), as reflected in Table~\ref{tab:metrics_combined_comparison}. Furthermore, the statistical analyses yield a Wilcoxon test p-value of \(1.16 \times 10^{-39}\), which confirms that the observed differences are statistically significant.

A key assumption underlying this system is that spoofing radar signals is highly challenging. For instance, an attacker attempting to spoof the AIS of a large vessel, such as an oil tanker, would need to generate a radar contact that matches the vessel's probabilistic signature. While it is theoretically possible to use an object of identical size to the intended AIS spoof, this approach offers minimal practical benefit and significantly increases the difficulty of successfully executing such an attack. Consequently, the DFCR system's confidence effectively penalises mismatched spoofed signals, enhancing the overall robustness of the system against adversarial spoofing attempts and outperforming baseline confidence models.

\section{Discussion} 
\label{sec:discuss}
This work aimed to address three critical challenges associated with adversarial AI: (1) the limited scope of traditional defences, (2) the inadequacy of current security metrics, and (3) the need for resilience that goes beyond model-based defences. To tackle these, we proposed developing AI defences with an approach (DFCR) to utilise multi-inputs and data fusion to create integrated defensive components.
 
The DFCR system addresses Challenge 1 by demonstrating its capability to defend against a range of attacks while reducing the limitations of traditional defences, which often compromise input quality through methods like input sanitation or degrade model accuracy through adversarial training. Instead, the DFCR approach preserves the input quality while enhancing defence robustness, ensuring that essential information remains intact for decision-making. For Challenge 2, we derived a novel AI security metric from this system, enabling the integration of security assessments directly into the decision-making process and offering a standardised way to measure the system's resilience. Finally, for Challenge 3, the DFCR defence-in-depth strategy enhances system resilience by layering DFCR defences; even if the input sanitation defence is bypassed, the system is still capable of rejecting adversarial data through alternative validation checks, ultimately strengthening protection against adversarial attacks.

Although poison-based attacks were excluded from this study, it is plausible to infer that the DFCR system's resilience could mitigate such threats. Suppose an optical detection model was poisoned to misidentify a target (e.g., confusing a buoy for a tanker). In that case, the system should flag this as anomalous if radar signatures do not match the optical contact. This highlights a potential capability for mitigating data poisoning effects. Likewise, this multi-source approach and defensive components could aid in detecting adversarial patches that aim to obscure or alter object identification.

The evaluation included rigorous testing through real-world scenarios and a comprehensive quantitative analysis. We compared the DFCR approach against single-input models and models utilising existing state-of-the-art defences. We assessed its performance against a suite of common open-box and closed-box attacks, including adversarial image perturbations, patch attacks, and sensor spoofing. The results demonstrated substantial resilience improvements: up to a 35\% reduction in the loss for multi-source perturbation attacks, 100\% for adversarial patch attacks, and 100\% for spoofing attacks. Many attacks failed entirely, as indicated by a confidence of zero, meaning the system successfully rejected these adversarial inputs.

Unlike some traditional defences, which can reduce detection accuracy, the DFCR approach maintained high detection reliability, a critical factor for real-world, high-risk applications where environmental noise could lead to increased false positives or negatives. The DFCR system also overcame biases seen in other state-of-the-art defences, such as input compression (dependent on preset compression value), which tends to remove only small perturbations, degrade the quality of the input image and, in the case of adversarial training, reduce normal model detection accuracy. Instead, the DFCR system validated diverse inputs to remove both small and large perturbations, as shown in Figure~\ref{fig:exp3_boxplot} as it is focused on validating different diverse inputs to make decisions. The DFCR system also did not degrade the original model performance, unlike adversarial training, or the quality of the input data, unlike input compression defence. In contrast, if current adversarial defence limitations are adequate for the application, existing state-of-the-art adversarial defences could be used in combination with the system, which is likely to extract further accuracy and robustness improvements.

Beyond maritime autonomy, this approach holds promise for securing a range of high-risk applications. As dataset availability, software, and hardware continue to advance, this multi-input DFCR approach could be useful for future resilient AI systems. Similar to humans integrating diverse sensory inputs (e.g., spatial, temporal, visual, audio) for decision-making, AI systems could achieve greater resilience by incorporating varied data sources. This research underscores that single-input model object detection remains highly vulnerable to adversarial attacks, which has policy implications for critical infrastructure and high-risk domains. For such applications, we advocate for initially integrating AI to assist human operators, allowing for safer operations and establishing trust before full automation.

This work does have limitations. The DFCR system's reliance on greater computational resources compared to single-model defences could pose challenges for deployment on resource-constrained edge devices. During the evaluation, the baseline model achieved an average inference time of $ 9.83 \times 10^{-2} $ seconds, while the DFCR system recorded an average inference time of $ 2.784 \times 10^{-1} $ seconds over a $ 4.870 \times 10^2 $ seconds scenario. Although the DFCR system, implemented in Python and not yet optimised, performed adequately on the hardware used in this study, it is important to note that other implementations tailored to specific applications — such as aerial systems — may require alternative defence components (and optimisation), potentially resulting in a faster or slower performance than the system implemented in this study.

We do not aim to guarantee an ``un-hackable'' system, as no system can ever be completely immune to compromise. Instead, by utilising a range of defensive components, the goal is to make it so costly (in resources, time, money, effort, and sophistication) for attackers that it becomes economically unviable, reducing attacker interest and risk \cite{tam2018cyber}.  

Future work could explore the effects of this approach for AI on the edge. Additionally, while we demonstrated robust defence mechanisms, no defence is entirely foolproof; further research is needed to assess the DFCR approach's resilience against a broader array of attacks and diverse data sources for decision-making, such as accessibility attacks (although new defence components may need to be developed). Recent attacks also consider edge computing-based attacks and resource exhaustion-based attacks \cite{navaneet2024slowformer}. For instance, an attacker could overwhelm the model's/system's heavy processes, such as correlation or the feedforward process, by introducing numerous contacts to the screen, potentially causing the device to crash. However, this type of attack is not considered within the scope of this work but may be considered in future work.

\section{Conclusions} 
\label{sec:Conclusion}
This study advances the development of secure, resilient systems against adversarial AI. As AI becomes more integral to high-risk sectors, developing diverse multi-input defence mechanisms (DFCR), as proposed in this work, will be crucial in safeguarding cyber-physical, transportation systems against increasingly sophisticated adversarial threats.

\section*{Acknowledgements}
The authors would like to thank the University of Plymouth for the use of their autonomous fleet. The authors would also like to extend their gratitude to David Bowman and Charlie Kay for their support throughout the deployment process.

\bibliographystyle{IEEEtran}
\bibliography{main}


 





\end{document}